\documentstyle[preprint,pra,aps,epsfig]{revtex}
\begin{document}
\tighten
\title{Universal quantum limits on single channel information, entropy and heat flow}
\author{Miles P. Blencowe\cite{auth}}
\address{Department of Physics and Astronomy, Dartmouth College, 
Hanover, New Hampshire 03755, USA}
\author{Vincenzo Vitelli\cite{auth2}}
\address{The Blackett Laboratory, Imperial College, London SW7 2BZ, UK}
\date{\today}
\maketitle
\begin{abstract}
We show that the recently discovered universal upper bound on the 
thermal conductance of a single channel comprising 
particles obeying arbitrary fractional statistics is in fact a consequence of a 
more general universal upper bound, involving the averaged entropy and energy
currents of a single channel connecting heat reservoirs with arbitrary 
temperatures and chemical potentials. The latter upper bound in turn leads, 
via Holevo's theorem, to a universal (i.e., statistics independent) 
upper bound on the  optimum 
capacity for classical information transmission down a 
single, wideband quantum channel.       
\end{abstract}


\section{Introduction}
\label{sec:intro}
In a recent experiment \cite{schwab}, Schwab {\it et al.} succeeded in 
measuring for the first time the thermal conductance quantum for 
a suspended, dielectric wire of submicron cross section. In accordance 
with predictions \cite{rego1}, the thermal 
conductance was found to approach the limiting value $4\times\pi 
k_{B}^{2}T/6\hbar\approx 4\times 10^{-12} T~\text{W\ K}^{-1}$ as the wire 
thermal reservoirs  were cooled such that the dominant phonon wavelength 
became comparable to the wire cross section. The factor of 4 is just 
the number of independent vibrational mode branches of the wire satisfying 
$\omega(k)\rightarrow 0$ as $k\rightarrow 0$ (see, e.g., Ref. 
\onlinecite{nishiguchi}). Only such modes can have non-negligible 
phonon occupation numbers as $T\rightarrow 0$, giving four available 
channels for heat transport. The single channel thermal conductance can never 
exceed the thermal conductance quantum $\pi k_{B}^{2}T/6\hbar$. 
The conductance quantum can only be attained for ballistic transport (i.e., no
scattering) as was achieved in the experiment. 

In common with the quantum limits for other single channel, linear transport 
coefficients,  such as the electronic conductance  quantum 
$e^{2}/h$ \cite{wees}, the thermal conductance 
quantum does not depend on the form of the $\omega(k)$ dispersion 
relations, a consequence of the cancellation of the group velocity  
and density of states factors in the formula for the one dimensional heat 
current. Wires made from different insulating materials and with 
different cross section geometries will therefore all have the same 
limiting single channel thermal conductance value for ballistic transport at 
low temperatures. For this reason, the conductance quantum is often 
termed `universal'. 

The thermal conductance is in fact universal in a much wider 
sense. For a single channel connecting two heat reservoirs 
with (quasi)particles obeying fractional  statistics according to 
Haldane's definition (which generalizes Bose and 
Fermi statistics) \cite{haldane}, it was recently found that the 
maximum, limiting thermal conductance quantum is independent of the 
particle statistics as 
well \cite{rego2,krive}. For example, in the case of an ideal electron gas, 
the limiting  single 
channel thermal conductance coincides with the above 
thermal conductance quantum for phonons.   While dimensional analysis 
would lead us to expect the same factor  $k_{B}^{2}T/\hbar$ to occur 
independently of the statistics, there is no a priori reason to 
expect the same numerical factor $\pi/6$ as well, given that the 
latter results from integrating with respect to the  energy the  
expansion to first order in  
small temperature differences  of the  thermal reservoir 
distributions, 
which have qualitatively different 
 forms for particles obeying different statistics. This remarkable property 
is unique to the thermal conductance: all other single channel 
transport coefficients depend on the particle statistics.

In an earlier and  unrelated investigation concerning the quantum 
limits on 
single channel information and entropy flow \cite{pendry1}, 
Pendry showed that the bound 
$\dot{S}^{2}/\dot{E}\leq\pi k_{B}^{2}/3\hbar$, involving the  averaged 
single channel entropy and energy currents, is obeyed for both  bosons and 
fermions. The striking resemblance between this bound and that for 
the single channel thermal conductance suggests the possible
existence of an universal and more general, attainable bound relating the entropy 
and energy currents, from which the  thermal conductance 
 bound would follow as a special case. In particular, there is the 
possibility of  
a bound which would be independent of the channel materials properties and 
particle statistics
and which would apply even far from equilibrium where the temperatures 
(and perhaps also the chemical 
potentials) of the two heat reservoirs connecting the ends of the channel
are significantly different.

Given that entropy and information are closely related, 
the existence of such a universal upper bound on the entropy flow rate 
 would in turn suggest the existence of an   
optimum capacity for single channel classical information 
transmission, 
which is also universal 
in the wider sense (i.e., independent of channel materials properties and 
particle 
statistics). This is in fact the main subject of Ref.\ 
\onlinecite{pendry1}. However, there the analysis is restricted to 
situations in which the channel is noiseless, with the information  encoded 
and decoded in terms of the  boson/fermion number eigenstates.
A proper determination of the optimum 
capacity would consider all possible input quantum states for encoding 
letters and all possible detection  schemes  at the output. The crucial 
result which allows one to generalize the analysis of Ref.\ 
\onlinecite{pendry1}  is Holevo's theorem \cite{holevo}, which bounds 
the mutual information between channel output and input with a 
quantity involving the quantum entropies of the input states. Caves 
and Drummond \cite{caves} have carried out the more general analysis 
for particles obeying Bose statistics only and confirm Pendry's upper 
bound as the optimum channel capacity.

Thus, there is the possibility of an optimum, universal limiting 
capacity which bounds all possible methods of encoding and detection, 
and which is independent of the physical properties of the channel. 
We emphasize that the existence of such a single channel 
optimum 
capacity is suggested by the established existence of the  universal 
thermal conductance quantum.

The fact that the thermal conductance by its very definition requires 
that the channel be connected at each end to a heat reservoir which 
can act both as an emitter and absorber of quanta, suggests that in 
order to attain the 
above-conjectured optimum capacity,  a generalization of 
the mutual information will be required in which there is a sender/receiver pair
at both ends of the channel. The pairs thus share the same channel 
and information can now flow in opposite directions.
In fact, as we shall see, with the exception of bosons the 
attainability of Pendry's upper bound for 
particles obeying arbitrary statistics necessarily 
requires that the chemical potentials of the two reservoirs coincide 
and be nonzero, so that both reservoirs are sources of particles.

In the following sections, we provide evidence for the validity 
of the two conjectures outlined above: namely (1) the existence of an 
universal bound relating the entropy and energy flow 
rates of a single quantum channel, with the universal thermal 
conductance bound following as a special case, and (2) the existence of an 
 universal bound on the optimum capacity for single-channel information 
transmission, subject to certain constraints on the channel and input 
states.  

In Sec. 
\ref{sec:entropy}, we first show that Pendry's bound holds for particles 
obeying arbitrary statistics under quite general conditions for 
the reservoir temperatures and chemical potentials. We then introduce a 
less general but tighter bound, which requires that the chemical 
potentials of the two reservoirs coincide, and replaces the energy 
current with the heat current $\dot{E} -\mu\dot{N}$. We show that the 
thermal conductance bound follows as a special case from this latter 
bound when the temperature difference between the two reservoirs 
approaches zero.

In Sec. \ref{sec:information}, we first repeat the analysis of Caves and 
Drummond \cite{caves} in the conventional case for unidirectional 
information transmission, but generalizing to particles obeying fractional 
statistics, and obtain the limiting statistics-dependent optimum capacity. We 
then generalize the definition of mutual information and Holevo's 
theorem in a simple way to allow for non-interfering, two-way 
information flow. Using this generalized definition we show, subject 
to certain constraints on the channel and input states, 
that the limiting optimum capacity is now 
independent of the particle statistics and coincides with that of the bosonic 
case. In the final part of the section, we introduce a further 
 generalization of the  mutual  information which 
allows for the possibility of interference between the `left-moving' and 
`right-moving' information flows.

In Sec. \ref{sec:conclusion}, we conclude and also briefly outline various open 
problems which have a bearing on the two conjectures.

\section{entropy bounds}
\label{sec:entropy}

As our generic model structure, we consider some confining `wire' which 
supports particles obeying a given statistics and 
which is connected adiabatically at each end to two particle reservoirs 
characterized by temperatures $T_{L}$ and $T_{R}$ and chemical potentials 
$\mu_{L}$ and $\mu_{R}$, where the subscripts $L$ and $R$ denote the left 
and right reservoirs, respectively. The device of Ref.\ \onlinecite{schwab} is 
one possible realisation of the model structure in the case of 
phonons. Typically, a wire will provide several available parallel channels 
for given reservoir chemical potential and temperature values. However, 
in the case of ballistic transport the channel currents don't interfere with 
each other and thus can be treated independently. We will restrict ourselves to 
ballistic transport in the present investigation. 

The distribution function for particles obeying fractional statistics 
is \cite{wu}
\begin{equation}
    f_{g}(E)=
    \left[w\left(\frac{E-\mu}{k_{B}T}\right)+g\right]^{-1},
    \label{dist1}
\end{equation}
where the function $w(x)$ satisfies
\begin{equation}
    w(x)^{g}[1+w(x)]^{1-g}=e^{x}.
    \label{dist2}
\end{equation}
The parameter $g$, assumed to be a rational number, determines the 
statistics. From these equations, we can see immediately that $g=0$ 
describes bosons and $g=1$ fermions.   The left(right) components of 
the single channel energy and entropy currents are \cite{rego2,krive}
\begin{equation}
    \dot{E}_{L(R)}=\frac{(k_{B}T_{L(R)})^{2}}{2\pi\hbar}
    \int_{x^{0}_{L(R)}}^{\infty}dx \left(x+\mu_{L(R)}/k_{B}T_{L(R)}
    \right) f_{g}(x)
    \label{energy}
\end{equation}
and
\begin{eqnarray}
    \dot{S}_{L(R)}=&-&\frac{k_{B}^{2}T_{L(R)}}{2\pi\hbar}
    \int_{x^{0}_{L(R)}}^{\infty}dx
    \left \{ f_{g}\ln f_{g}+(1-g f_{g}) \ln(1-g f_{g})\right.\cr
    &-&\left. [1+(1-g) f_{g}] 
    \ln [1+(1-g) f_{g}]\right \},
    \label{entropy}
\end{eqnarray}
where $x^{0}_{L(R)}=-\mu_{L(R)}/k_{B}T_{L(R)}$, 
and where we define the energy origin such that the minimum energy of a
channel particle is zero, i.e., the energy is given by the 
longitudinal kinetic component. The total energy and entropy 
channel currents are then just 
$\dot{E}=\dot{E}_{L}-\dot{E}_{R}$ and 
$\dot{S}=\dot{S}_{L}-\dot{S}_{R}$, respectively.

The first conjectured bound involving these single channel entropy and energy 
currents is \cite{pendry1}
\begin{equation}
   \dot{S}^{2}\leq \frac{\pi k_{B}^{2}}{3\hbar}\dot{E},
    \label{bound1}
\end{equation}
provided $T_{L}>T_{R}$ and $\mu_{L}\geq\mu_{R}$. We have numerically tested this 
bound extensively in $\mu$ and $T$ parameter space, for
several rational values of the 
statistical parameter $g$ ranging between between zero and one. Fig.~1 
gives an initial idea of the bound  
by showing the dependence of the ratio 
${3\hbar}\dot{S}^{2}/\pi k_{B}^{2}\dot{E}$ on a selected parameter 
range. 

In the case of bosons with constant $\mu_{L}=\mu_{R}=0$ (e.g., photons 
or phonons), 
evaluating Eqs. (\ref{energy}) and (\ref{entropy}) gives 
$\dot{E}_{L(R)}=\pi(k_{B}T_{L(R)})^{2}/12\hbar$ and 
$\dot{S}_{L(R)}=\pi k_{B}^{2}T_{L(R)}/6\hbar$, 
respectively. Setting $T_{R}=0$ and eliminating $T_{L}$ by solving 
for $\dot{S}(=\dot{S}_{L})$ in terms of $\dot{E}(=\dot{E}_{L})$, we 
obtain equality in bound (\ref{bound1}). 
For all other physically  achievable parameter choices, we have strict inequality in
(\ref{bound1}). The key point, however, is that the bound can be 
approached arbitrarily 
closely, no matter the particle statistics. For example, in the case 
of bosons with non-constant reservoir chemical potentials, the bound 
is approached asymptotically in the  degenerate limit  
$-x^{0}_{L}=\mu_{L}/k_{B}T_{L}\rightarrow 
0^{-}$, with $\mu_{R}=0$ and $T_{R}=0$.
For particles with $g>0$, the bound is approached asymptotically in 
the degenerate limit
$-x^{0}_{L}=\mu_{L}/k_{B}T_{L}\rightarrow +\infty$, with 
$\mu_{L}=\mu_{R}$ and $T_{R}=0$. That these are the correct conditions 
for approaching the upper bound can be seen more clearly after 
transforming the integrals in Eqs. (\ref{energy}) and (\ref{entropy}) 
for the bosonic and fermionic cases as in, e.g., Sec. 58 of 
Ref.\ \onlinecite{landau}. For example, in 
the case of fermions,  the single channel  energy and entropy currents can be 
rewritten as follows
\begin{eqnarray}
    \dot{E}=\frac{\pi (k_{B}T_{L})^{2}}{12 \hbar}\Bigg[ 
    1-\left(\frac{T_{R}}{T_{L}}\right)^{2} &+&\frac{3}{(\pi 
    k_{B}T_{L})^{2}}\left(\mu_{L}^{2}-\mu_{R}^{2}\right) 
    +\frac{6}{\pi^{2}}\int_{-x^{0}_{L}}^{\infty} 
    dx\left(\mu_{L}/k_{B}T_{L} -x\right) f(x)\cr &-&
    \frac{6}{\pi^{2}}\left(\frac{T_{R}}{T_{L}}\right)^{2}
    \int_{-x^{0}_{R}}^{\infty} 
    dx\left(\mu_{R}/k_{B}T_{R} -x\right) f(x)\Biggr]
    \label{energy2}
\end{eqnarray}
and
\begin{eqnarray}
    \dot{S}=\frac{\pi k_{B}^{2}T_{L}}{6 \hbar}
    \Biggl\{1-\frac{T_{R}}{T_{L}} &+& 
    \frac{3}{\pi^{2}}\int_{-x^{0}_{L}}^{\infty} 
    dx\left[f\ln f +(1-f)\ln (1-f)\right]\cr &-&
    \frac{3}{\pi^{2}}\frac{T_{R}}{T_{L}}
    \int_{-x^{0}_{R}}^{\infty} 
    dx\left[f\ln f +(1-f)\ln (1-f)\right]\Biggr\}.
    \label{entropy2}
\end{eqnarray}
Taking the 
limit $-x^{0}_{L}\rightarrow +\infty$, with the conditions $\mu_{L}=\mu_{R}$ 
and $T_{R}=0$, only the first term remains on 
the right-hand-sides of Eqs. (\ref{energy2}) and (\ref{entropy2}) and 
the energy and entropy currents coincide with those for bosons with 
$\mu_{L}=\mu_{R}=0$ and $T_{R}=0$.

It is not possible to recover the single channel thermal conductance 
bound from (\ref{bound1}). The best we can do is to derive an 
upper bound on the rate of heat emission from an isolated reservoir 
for bosons with zero chemical potential \cite{pendry1} (see also Ref.\ 
\onlinecite{pendry2}). Setting $\mu_{L}=\mu_{R}=0$, $T_{R}=0$, identifying 
the heat emission rate with the total energy emission 
rate  $\dot{Q}_{L}=\dot{E}_{L}$, and using $\dot{Q}_{L}/T_{L}\leq\dot{S}_{L}$, 
bound (\ref{bound1}) gives 
\begin{equation}
    \dot{Q}_{L}\leq \frac{\pi k_{B}^{2}T_{L}^{2}}{3\hbar}.
    \label{heat}
\end{equation}
Note that for particles with nonzero chemical potential, the heat 
emission 
rate is $\dot{Q}_{L}=\dot{E}_{L}-\mu_{L} \dot{N}_{L}$, where 
$\dot{N}$ denotes the number current, and in this case
(\ref{heat}) does not follow from (\ref{bound1}). If we had equality 
in (\ref{heat}), then the thermal conductance could be 
obtained by taking the difference $\dot{Q}_{L}-\dot{Q}_{R}=\pi 
k_{B}^{2} (T_{L}^{2}-T_{R}^{2})/3\hbar =2\pi k_{B}^{2}\bar{T}\delta 
T/3\hbar$, where $\bar{T}=(T_{L}+T_{R})/2$. But this gives the 
incorrect coefficient ($2/3$ instead of 
$1/6$). What is wrong with this argument is the assumption that 
$\dot{Q}_{L(R)}/T_{L(R)}=\dot{S}_{L(R)}$. In fact, 
$\dot{Q}_{L(R)}/T_{L(R)}=\dot{S}_{L(R)}/2$, signalling the 
irreversible nature of the heat emission.

A conjectured, tighter bound suggested by the form of expressions 
(\ref{energy2}) and (\ref{entropy2}) which does 
yield the thermal conductance bound as a special case, is the following
\begin{equation}
    \dot{S}^{2}\leq \frac{\pi k_{B}^{2}}{3\hbar}\left(\frac{T_{L}-T_{R}}
    {T_{L}+T_{R}}\right)\left(\dot{E}-\mu\dot{N}\right),
    \label{bound2}
\end{equation}
provided $T_{L}>T_{R}$ and $\mu_{L}=\mu_{R}=\mu$. 
Again, we have numerically tested this 
bound extensively in $\mu$ and $T$ parameter space, for
several rational values of the 
statistical parameter $g$ ranging between between zero and one (see 
Figs. 2 and 3).
In the case of bosons with constant $\mu=0$, we 
obtain equality for all $T_{L}>T_{R}$. 
For  bosons with non-constant $\mu\leq 0$, the bound 
is approached asymptotically in the  degenerate limit  
 $\mu/k_{B}T_{L(R)}\rightarrow 
0^{-}$.
For particles with $g>0$, the bound is approached asymptotically in 
the degenerate limit
$\mu/k_{B}T_{L(R)}\rightarrow +\infty$. Note that the heat 
current $\dot{Q}=\dot{E}-\mu \dot{N}$ appears instead of the energy current
$\dot{E}$ on the right-hand-side of bound (\ref{bound2}).  
This  replacement is essential: if the energy 
current is used, then the bound can be violated for bosons with  
$\mu<0$. It is remarkable that the need to recover the thermal 
conductance and also to satisfy the bound both lead to the replacement 
of the energy current with the heat current.   

\section{information bounds}
\label{sec:information}

Consider a communication channel, characterised by an input alphabet 
$A$ with letters labeled by an index $a=1,\dots ,{\cal A}$ and a set 
of probabilities $p_{A}(a)$ for transmitting letter $a$, an output 
alphabet $B$ labeled by $b=1,\dots,{\cal B}$, and a set of conditional 
probabilities $p_{B|A}(b|a)$ for receiving letter $b$, given transmission of 
letter $a$. The mutual information gives the measure of the 
information successfully transmitted from input to output of the communication 
channel:
\begin{equation}
    H(B;A)=\sum_{a,b}p_{B|A}(b|a)p_{A}(a) \log_{2}\left(\frac{p_{B|A}(b|a)}
    {p_{B}(b)}\right),
    \label{mutual}
\end{equation}
where $p_{B}(b)=\sum_{a}p_{B|A}(b|a) p_{A}(a)$. 

Suppose the quantum channel 
medium supports particles for some given rational, 
statistical parameter value $g$,  $0\leq 
g\leq 1$. Let the input 
letter $a$ be  encoded in some quantum state $\hat{\rho}_{a}$, and 
the output detection scheme be described, in the most general case, by a set 
of non-negative, bounded Hermitian operators $\hat{F}_{b}$ satisfying 
$\sum_{b}\hat{F}_{b}=\hat{1}$, with 
$p_{B|A}(b|a)=\text{tr}(\hat{\rho}_{a}\hat{F}_{b})$. The operators 
$\hat{\rho}_{a}$ and $\hat{F}_{b}$ act on the channel Fock space 
for statistical parameter value $g$. More precisely, since information is 
transmitted in only one direction,  these 
operators act on the subspace describing  right-moving states, say.

Holevo's theorem \cite{holevo} provides an upper bound on the mutual 
information for all possible detection schemes:
\begin{equation}
    \max_{\{\hat{F}_{b}\}} H(B;A)\leq S(\hat{\rho})-\sum_{a}p_{A}(a) 
    S(\hat{\rho}_{a}),
    \label{holevo}
\end{equation}
where $\hat{\rho}=\sum_{a} p_{A}(a)\hat{\rho}_{a}$ and 
$S(\hat{\rho})=-{\text tr}(\hat{\rho}\log_{2}\hat{\rho})$ is the 
quantum entropy in bits. Note that, while this theorem is usually 
applied to bosonic communication channels (c.f., Ref.\ 
\onlinecite{caves}), it is in fact applicable to
channels for arbitrary fractional statistics. All that is required 
is that the channel obeys the usual rules of 
quantum mechanics.

Maximizing the mutual information  with respect to the output detection 
scheme and the input states and probabilities gives the 
optimum capacity $C$ of the channel. From (\ref{holevo}), we have 
\cite{caves}
\begin{equation}
    C=\frac{1}{\cal 
    T}\max_{\hat{\rho}}\max_{\{\hat{F}_{b}\}}H(B;A)\leq
    \frac{1}{\cal 
    T}\max_{\hat{\rho}} S(\hat{\rho})=\frac{S_{\text {max}}}{\cal T},
    \label{capacity}
\end{equation}
where $\cal{T}$ is the transmission time. As shown in Sec. IV.B of Ref. 
\onlinecite{caves}, the upper bound $S_{\text {max}}$ can in fact be 
attained: Find a complete, orthonormal set of diagonalizing basis 
states $|a\rangle$ for the $\hat{\rho}$ which maximizes $S(\hat{\rho})$, 
i.e., $\hat{\rho}=\sum_{a}q(a)|a\rangle\langle a|$. Choose 
$\hat{\rho}_{a}=|a\rangle\langle a|=\hat{F}_{a}$ and $p_{A}(a)=q(a)$. 
Then $H(B;A)=-\sum_{a}p_{A}(a)\log_{2}p_{A}(a)=S(\hat{\rho})=S_{\text{max}}$. 

Thus, the optimum capacity is just the maximum quantum entropy in bits 
divided by the  transmission time, subject
to the given constraints on the channel. One common constraint is to fix the 
total energy of the transmitted message; the optimum capacity then 
gives the maximum information that can be transmitted in a time 
${\cal T}$ for a given, 
allowed signal energy. For a single, wideband channel with 
longitudinal single-particle energies $h f_{j} =h j/{\cal T}$, $j=1,2,\dots$, 
the total longitudinal energy $E_{N}$ of a given Fock state is 
$E_{N}=\sum_{j} hf_{j} 
n_{j} ={Nh}/{\cal T}$, where $N=\sum_{j=1}^{\infty}j n_{j}$, and $n_{j}$ 
is the occupation number of, say, the right-propagating mode $j$. The 
maximum entropy is then $S_{\text{max}}=\log_{2}{\cal N}_{N}$, 
where ${\cal N}_{N}$ is just the number of different ways the sum $N$ can be 
partitioned. For bosons ($g=0$), ${\cal N}_{N}$ is given by 
the number of unrestricted partitions, while for the fermions 
($g=1$), ${\cal 
N}_{N}$ is given by the number of partitions into distinct parts. More 
generally, for particles obeying fractional statistics with $g=1/n$, 
$n=1,2,\dots$, no number can appear more than $n$ times in a given 
partition. Note, however, that for $0<g<1$ there are additional 
constraints on the allowed partitions, as discussed in Ref.\ 
\onlinecite{murthy}. 

For long transmission times  ${\cal T}$, or equivalently large $N$, 
one obtains  the following asymptotic approximation to the optimum 
capacity of a wideband, bosonic channel for fixed energy \cite{caves}:
\begin{equation}
    C_{\text{boson}}=\frac{\pi}{\ln 2}\sqrt{\frac{2 P}{3 h}}-\frac{1}{\cal T} 
    \log_{2}\left(\frac{4\sqrt{3} P {\cal T}^{2}}{h}\right),
    \label{bosecapacity}
\end{equation}
where $P=E_{N}/{\cal T}$ is the time-averaged power. Caves {\it et 
al.} \cite{caves}
also derive the bosonic optimum  capacity subject to the  alternative 
constraints that the maximum-energy or the message-ensemble-averaged-energy 
of the channel be fixed. All give the same leading order term as on the 
right-hand-side of Eq. (\ref{bosecapacity}), with $P$ appropriately defined in 
each case. 

In order to write down the long transmission-time optimum capacity of a 
wideband, fermionic channel for fixed energy, we require the asymptotic 
approximation to the number of distinct partitions of $N$ 
(see, e.g., Sec. 24.2.2 of Ref. \cite{abramowitz}):
\begin{equation}
    {\cal N}_{N}\sim \frac{1}{4\cdot 3^{1/4}\cdot 
    N^{3/4}}e^{\pi\sqrt{1/3}\sqrt{N}}.
    \label{distinct}
\end{equation}
This gives 
\begin{equation}
    C_{\text{fermion}}=\frac{\pi}{\ln 2}\sqrt{\frac{P}{3 h}}-\frac{1}{\cal T} 
    \log_{2}\left[4\cdot 3^{1/4} \left(\frac{P {\cal 
    T}^{2}}{h}\right)^{3/4}\right].
    \label{fermicapacity}
\end{equation}
Note that, in the limit ${\cal T}\rightarrow\infty$, the fermionic 
optimum capacity is smaller than the bosonic optimum capacity by a 
factor $\sqrt{2}$ for given power $P$. Note also that these optimum 
capacities  satisfy the information-theoretic 
counterpart to bound (\ref{bound1}) \cite{pendry1}:
\begin{equation}
    C< \frac{\pi}{\ln 2}\sqrt{\frac{2 P}{3 h}}
    \label{infobound}
\end{equation}
for finite ${\cal T}$.

Asymptotic approximations to $C$, 
analogous to  Eqs. (\ref{bosecapacity}) and 
(\ref{fermicapacity}),  can no doubt also be written down for certain other 
rational $g$ values. However, rather than attempting to derive $C$ 
through the nontrivial route which involves first obtaining the 
asymptotic approximation to the number of partitions, we can appeal 
to the fact that different ensemble derivations of the entropy
give the same result in the thermodynamic limit when the ensemble energies 
coincide. In 
particular, we can 
instead use expressions (\ref{energy}) and (\ref{entropy}) for  
$\mu=0$  (only the energy is constrained and not the particle 
number) to derive 
the leading order term in the asymptotic approximation to $C$. For 
example, in the case of `semions' ($g=1/2$), carrying 
out the integrals in (\ref{energy}) and (\ref{entropy}) 
 gives 
$C_{\text{semion}}({\cal T\rightarrow\infty})=(\pi/\ln 2)\sqrt{2 
P/5h}$, 
which falls between the Fermi and Bose 
capacities,  again satisfying  bound (\ref{infobound}).  Solving 
numerically Eqs. (\ref{energy}) and (\ref{entropy}) for a range of 
$g$-values, we find that 
$C_{g}({\cal T\rightarrow\infty})$ decreases monotonically as $g$ 
increases from 0 to 1 (Fig.~4).  Thus bound (\ref{infobound}) holds for all 
$0\leq g\leq 1$.

However, unlike the analogous upper bound (\ref{bound1}) on the single-channel 
physical entropy current, the information-theoretic bound (\ref{infobound})
cannot be approached arbitrarily closely independently of the particle 
statistics: only for bosons is the upper bound approached in the limit
${\cal T}\rightarrow\infty$. 
But recall, from the form of the conditions for approaching the upper bound 
(\ref{bound1}), and also from the form of Eqs. (\ref{energy2}) and 
(\ref{entropy2}) for fermions, that it is crucial for both ends of the channel to be 
connected to reservoirs providing two-way energy and entropy flows in the 
channel. This suggests that, with a suitable generalization of the 
communication channel allowing for two-way information flow, the 
channel capacity will approach the upper bound (\ref{infobound}) 
arbitrarily closely independently of the particle statistics.

Consider, therefore, two sender-receiver `stations', with one station at each end of 
the single channel, thus sharing the channel. 
Station $L$ at the left end encodes information in right-moving states and 
decodes 
information from left-moving states, while station $R$ at the right 
end encodes 
information in left-moving states and decodes information from 
right-moving states. A single `use' of the channel involves $L$ and 
$R$ each sending and subsequently detecting a message, the whole 
operation taking place during an interval ${\cal T}$. Station $L$ 
uses an input alphabet $A_{L}$ with letters labeled by an index 
$a_{L}=1,\dots ,{\cal A}_{L}$ and a set 
of probabilities $p_{A_{L}}(a_{L})$ for transmitting letter $a_{L}$, 
and an output 
alphabet $B_{L}$ labeled by $b_{L}=1,\dots,{\cal B}_{L}$. Station  $R$ 
similarly uses an input alphabet $A_{R}$ with transmission 
probabilities $p_{A_{R}}(a_{R})$, and an output alphabet $B_{R}$. 
The probability that $R$  receives letter $b_{R}$, given that $L$ 
sends letter $a_{L}$ is denoted as  $p_{B_{R}|A_{L}}(b_{R}|a_{L})$, and 
analogously  for the  conditional probability $p_{B_{L}|A_{R}}(b_{L}|a_{R})$.
We assume throughout that the joint probabilities for $L$ and $R$ to 
send a message are uncorrelated, i.e., $p_{A_{L},A_{R}}(a_{L},a_{R})=
p_{A_{L}}(a_{L}) p_{A_{R}}(a_{R})$.  
 We also assume to begin with that the left- and right-moving 
information flows do not interfere with each other.

Using the formula for the single channel entropy current as a guide 
[see, e.g., Eqs. (11) and (12) of Ref.\ \onlinecite{sivan}], we 
define the {\it net} information  transmitted from the $L$ 
and $R$ inputs to the $R$ output during a single use of the channel to 
be
\begin{equation}
    H(B_{R};A_{L},A_{R})=H(B_{R};A_{L})-H(A_{R}),
    \label{generalmutual}
\end{equation}
where $H(B_{R};A_{L})$ is defined as in Eq. (\ref{mutual}) and 
$H(A_{R})=-\sum_{a_{R}}p_{A_{R}}(a_{R})\log_{2}p_{A_{R}}(a_{R})$.  
Similarly, the 
net information  transmitted to the $L$ output is 
$H(B_{L};A_{R},A_{L})=H(B_{L};A_{R})-H(A_{L})$. Note the asymmetry 
of the two terms on the right-hand-side 
of definition (\ref{generalmutual}), reflecting an analogous 
asymmetry of the left- and right-moving components making up the 
net 
entropy current \cite{sivan}.  With the information defined 
with respect to the receiver at the  right end of the channel, 
it makes more sense to use the  information $H(A_{R})$ rather 
than the mutual information $H(B_{L};A_{R})$ which takes into account 
the channel noise and receiver properties at the other end of 
the channel. As we shall soon see when we generalize 
(\ref{generalmutual}) to include interfering left- 
and right-moving information, one reason why it might be a 
good thing to subtract, rather than to add, the information $H(A_{R})$ 
is that it gives reasonable answers in familiar examples such as that 
of a returned or `bounced' message.

But perhaps the most appealing property of $H(B_{R};A_{L},A_{R})$ as 
defined is that it satisfies a generalized Holevo theorem:
\begin{equation}
    \max_{\{\hat{F}_{b_{R}}\}}H(B_{R};A_{L},A_{R})\leq S(\hat{\rho}_{L}) -
    S(\hat{\rho}_{R}) 
    -\sum_{a_{L}}p_{A_{L}}(a_{L})S(\hat{\rho}_{a_{L}})+
    \sum_{a_{R}}p_{A_{R}}(a_{R})S(\hat{\rho}_{a_{R}}),
    \label{generalholevo}
\end{equation}
where equality holds if and only if the left input states 
$\hat{\rho}_{a_{L}}$ commute and the right input states $\hat{\rho}_{a_{R}}$ 
are orthogonal (see, e.g., Sec. IV.B of Ref.\ \onlinecite{caves}). 
Inequality (\ref{generalholevo}) is a consequence both of the
Holevo theorem (\ref{holevo}), which bounds $H(B_{R};A_{L})$,  
and also of the inequality $H(A_{R})\geq 
S(\hat{\rho}_{R})-\sum_{a_{R}}p_{A_{R}}(a_{R})S(\hat{\rho}_{a_{R}})$. 
Note that the latter inequality goes in the opposite direction to 
that of (\ref{holevo}), so that one must subtract $H(A_{R})$ in order 
that $H(B_{R};A_{L},A_{R})$ be bounded.

Maximizing the information $H(B_{R};A_{L},A_{R})$ with respect to the 
$R$ output detection scheme and the $L$ and $R$ input states and 
probabilities gives the optimum capacity of the channel. From 
(\ref{generalholevo}), we have
\begin{equation}
    C=\frac{1}{\cal 
    T}\max_{\hat{\rho}_{L},\hat{\rho}_{R}}\max_{\{\hat{F}_{b_{R}}\}}
    H(B_{R};A_{L},A_{R})\leq
    \frac{1}{\cal 
    T}\max_{\hat{\rho}_{L}} S(\hat{\rho}_{L})=\frac{S_{\text {max}}}{\cal T}.
    \label{generalcapacity}
\end{equation}    
The upper bound $S_{\text {max}}$ can in fact be 
attained: find a complete, orthonormal set of diagonalizing basis 
states $|a_{L}\rangle$ for the $\hat{\rho}_{L}$ which maximizes 
$S(\hat{\rho}_{L})$, 
i.e., $\hat{\rho}_{L}=\sum_{a_{L}}q(a_{L})|a_{L}\rangle\langle 
a_{L}|$. Choose 
$\hat{\rho}_{a_{L}}=|a_{L}\rangle\langle a_{L}|=\hat{F}_{a_{L}}$ 
and $p_{A_{L}}(a_{L})=q(a_{L})$. Choose any set 
$\{\hat{\rho}_{a_{R}}\}$ and probabilities 
$p_{A_{R}}(a_{R})=\delta_{a_{R},a_{R}'}$ for some fixed $a'_{R}$. 
Then $H(B_{R};A_{L},A_{R})=H(B_{R};A_{L})=-\sum_{a_{L}}p_{A_{L}}(a_{L})
\log_{2}p_{A_{L}}(a_{L})=S(\hat{\rho}_{L})=S_{\text{max}}$. The 
choice for $\hat{\rho}_{a_{R}}$ and $p_{A_{R}}(a_{R})$  
reflects the obvious fact that, for the  definition  (\ref{generalmutual}),
maximizing $H(B_{R};A_{L},A_{R})$ requires that $H(A_{R})$ be 
minimized, so that any left-moving message component can be sent, 
provided it is with probability one so that its information content is 
zero.  
 
Thus, the optimum capacity is just the maximum quantum entropy in bits for 
right-moving states divided by the  transmission time, subject 
to the given constraints on the channel. Of particular interest are 
the constraints for which the optimum capacity in the limit ${\cal 
T}\rightarrow\infty$ is independent of the statistical parameter $g$. 
Recalling the conditions for approaching asymptotically the entropy 
bound (\ref{bound1}) for $g>0$, 
namely $\mu_{L}/k_{B}T_{L}\rightarrow +\infty$ 
with $\mu_{L}=\mu_{R}$ and $T_{L}>T_{R}=0$, a little thought 
establishes that two  constraints are: fixed power (i.e., fixed energy 
current) $P>0$ and fixed number current $\dot{N}=0$. Again, as for the 
 unidirectional optimum capacity, the choice of 
ensemble for the definition of $P$ and 
$\dot{N}$---microcanonical, grand canonical etc.---is immaterial in the limit 
${\cal T}\rightarrow\infty$. Given that the unidirectional
optimum capacity for $0<g\leq 1$ is strictly less than the 
 bosonic optimum capacity 
$(\pi/\ln 2)\sqrt{2P/3h}$ in the limit ${\cal 
T}\rightarrow\infty$,     
it may seem paradoxical that  additional constraints 
have to be imposed (namely, $\dot{N}=0$) in order to attain the 
latter, larger capacity. The resolution lies in the fact that the dimension 
of the channel Hilbert space accessible for information and energy 
transmission has been doubled through the accomodation of left-moving 
states. 

The two above constraints, while necessary, are not sufficient. The 
problem lies in the fact that optimization step 
(\ref{generalcapacity}) places no conditions on the left-moving states  
$\hat{\rho}_{a_{R}}$, with the result that it is rather easy to find 
examples  where the power $P$ can be made arbitrarily small for 
given $S_{{\text{max}}}$, while at the same time satisfying the 
constraint $\dot{N}=0$. One possible way to overcome this problem is 
to introduce the further constraint on the left-moving states  
$\hat{\rho}_{a_{R}}$ that they be completely degenerate. 
Then it is possible to show that $S_{{\text{max}}}=(\pi/\ln 
2)\sqrt{2P/3h}$ in the limit ${\cal T}\rightarrow\infty$, 
i.e., $S_{{\text{max}}}$ coincides with the limiting, unidirectional   
bosonic optimum capacity independently of $0<g\leq 1$. Furthermore,
in the case of bosons
adding a left-moving degenerate state  does not change the energy 
current, so that  the above constraints can also be applied to bosons with the 
unidirectional bosonic optimum capacity again being obtained in the limit.   

What we have essentially done both here and in the previous section
is cancel  part of the 
right-moving energy current component  with a left-moving, degenerate component, 
leaving the information and entropy currents unchanged, thus increasing 
the  optimum capacity and  
entropy current bound  for given energy current [Eqs. 
(\ref{energy2}) and (\ref{entropy2}) show this more explicitly]. What is 
remarkable is that the  optimum capacity (\ref{infobound}) and 
entropy current bound
(\ref{bound1}) are 
attained asymptotically for a common set of constraints independent 
of the statistics  $0\leq g\leq 1$. 

In the final part of this section, we generalize our two-way 
information definition (\ref{generalmutual}) so as to allow for the 
possibility of interference between the left- and right-moving 
information flows. 
Our definition is motivated by the formula for 
the single-channel entropy current in the presence of elastic 
scattering in the channel \cite{sivan}. We define the {\it net} 
information  transmitted from the $L$ 
and $R$ inputs to the $R$ output during a single use of the channel to 
be
\begin{eqnarray}
    H(B_{R};A_{L},A_{R})&=&\sum_{a_{L},a_{R},b_{R}}p_{B_{R}|A_{L},A_{R}}
    (b_{R}|a_{L},a_{R})\ p_{A_{L}}(a_{L})\ p_{A_{R}}(a_{R})\cr&\times&
    \log_{2}\left({p_{B_{R}|A_{L},A_{R}}
    (b_{R}|a_{L},a_{R})}/{p_{B_{R}}(b_{R})}\right)+
    \sum_{a_{R}}p_{A_{R}}(a_{R})\log_{2}p_{A_{R}}(a_{R})
    \label{generalmutual2},
\end{eqnarray}
with an analogous definition for $H(B_{L};A_{R},A_{L})$ and where we again
assume that the joint probabilities for $L$ and $R$ to 
send a message are uncorrelated.  The channel interference is 
conveniently implemented by a unitary `scattering' operator $\hat{{\cal S}}$, 
acting on the states as 
$\hat{{\cal S}}(\hat{\rho}_{L}\otimes\hat{\rho}_{R})
\hat{{\cal S}}^{\dag}\rightarrow
\hat{\rho}'_{L}\otimes\hat{\rho}'_{R}$, where we restrict ourselves to 
non-correlating interfering processes. The conditional
probabilities are constructed as follows:
\begin{mathletters}
    \label{allequation}
\begin{equation}
    p_{B_{R}|A_{L},A_{R}}(b_{R}|a_{L},a_{R})={\text{tr}}\left[
    (\hat{F}_{b_{R}}\otimes\hat{1})
    \hat{{\cal S}}(\hat{\rho}_{a_{L}}
    \otimes\hat{\rho}_{a_{R}})\hat{{\cal S}}^{\dag}\right]
    \label{generalconditionala}
\end{equation}
and
\begin{equation}
    p_{B_{L}|A_{L},A_{R}}(b_{L}|a_{L},a_{R})={\text{tr}}\left[
    (\hat{1}\otimes\hat{F}_{b_{L}})
    \hat{{\cal S}}(\hat{\rho}_{a_{L}}\otimes\hat{\rho}_{a_{R}})
    \hat{{\cal S}}^{\dag}\right]
    \label{generalconditionalb}, 
\end{equation}
\end{mathletters}
where the the right and left  detector operators are written as 
$\hat{F}_{b_{R}}\otimes\hat{1}$ and $\hat{1}\otimes\hat{F}_{b_{L}}$, 
respectively, 
reflecting the fact that, in the absence of interference, 
i.e., when $\hat{{\cal S}}$ is 
the identity operator, the 
right(left) detector can only receive left(right) input states.
Note that definition (\ref{generalmutual2}) reduces
to the two-way information definition (\ref{generalmutual}) when 
there is no interference.

The more general, two-way information definition (\ref{generalmutual2}) can be 
applied to certain situations which are beyond the scope of the 
unidirectional mutual information (\ref{mutual}). 
As a simple  example, consider the situation of a `bounced' message, 
an all too common occurence with electronic mail.  This example can be 
modeled as follows: let the right letters be encoded in the orthonormal states 
$|a_{R}\rangle$ and sent with probabilities $p_{A_{R}}(a_{R})$. Let 
the states $|a_{L}\rangle$  encoding the left letters, and sent with 
probabilities $p_{A_{L}}(a_{L})$,  be in 
one-to-one correspondence with the right states $|a_{R}\rangle$, 
with the mapping achieved simply by reversing the propagation 
direction. Let the right detector be characterized by 
projection operators $\hat{F}_{a_{R}}=|a_{L}\rangle\langle a_{L}|$.   
Finally, suppose the scattering operator reverses the direction of the 
propagating states, i.e., $\hat{{\cal S}} |a_{R(L)}\rangle = 
|a_{L(R)}\rangle$. Then evaluating the two-way information 
(\ref{generalmutual2}), we find that the first term on the 
right-hand-side reduces to the information $H(A_{R})$, thus cancelling the 
second term and giving the value $H(B_{R};A_{L},A_{R})=0$. This 
coincides with our common-sense measure: if a message gets bounced 
back, then no information was sent.

From the  Holevo theorem (\ref{holevo}) for unidirectional information 
flow and also the 
inequality $H(A_{R})\geq 
S(\hat{\rho}_{R})-\sum_{a_{R}}p_{A_{R}}(a_{R})S(\hat{\rho}_{a_{R}})$,
it is straightforward to show that the information 
$H(B_{R};A_{L},A_{R})$ as defined in (\ref{generalmutual2}) also satisfies 
a generalized Holevo theorem:
\begin{equation}
   \max_{\{\hat{F}_{b_{R}}\},\hat{\cal S}}H(B_{R};A_{L},A_{R})\leq 
   S(\hat{\rho}_{L})-\sum_{a_{L}}p_{A_{L}}(a_{L})S(\hat{\rho}_{a_{L}})
   \label{generalholevo2}.
\end{equation}    
Such a bound enables us to determine the optimum capacity allowing 
also for interfering left- and right-moving information flows. 
We shall leave this to a future investigation.

\section{conclusion}
\label{sec:conclusion}

We have provided evidence for the validity of two related conjectures 
which state that the entropy current and optimum capacity 
for information transmission of a single channel are universally 
bounded for given energy current/power, independently of the channel 
materials properties and particle statistics according to Haldane's 
definition.  What is most 
notable, is that these bounds can be approached arbitrarily closely 
no matter the particle statistics. A less general, tighter bound on 
the entropy current was also conjectured, from 
which the recently discovered statistics-independent thermal 
conductance bound follows as a special case. The
statistics-independent, limiting bound on the optimum capacity 
required a generalisation of the definition for the transmitted 
information, allowing for two-way information flow. The bound then 
followed from a generalized Holevo theorem, with certain constraints placed 
on the channel and input states.

The results presented here can be extended in several ways. It would 
be more satisfying to have an analytic proof of the conjectured bounds, 
rather than an exhaustive numerical check. The entropy current bound 
should  be tested under more general conditions, for example in the 
presence of channel scattering. Similarly, the optimum capacity bound 
should be tested also allowing for interference between two-way 
information flows. 

Finally, we point out the recent demonstration that  Holevo's theorem 
follows from Landauer's principle of information erasure \cite{plenio}. 
In the light of this, it would be interesting to try to rederive the 
universal upper bound on the optimum capacity starting from 
Landauer's erasure principle.

\acknowledgements
M.P.B. thanks Michael Roukes and his group at Caltech for stimulating 
discussions and for their hospitality during a visit. 
We  would also like to thank Jay Lawrence and Martin Plenio for 
helpful discussions and also for suggesting improvements to the 
manuscript. V.V. was partially funded by a Nuffield Foundation Bursary for 
Undergraduate Research.

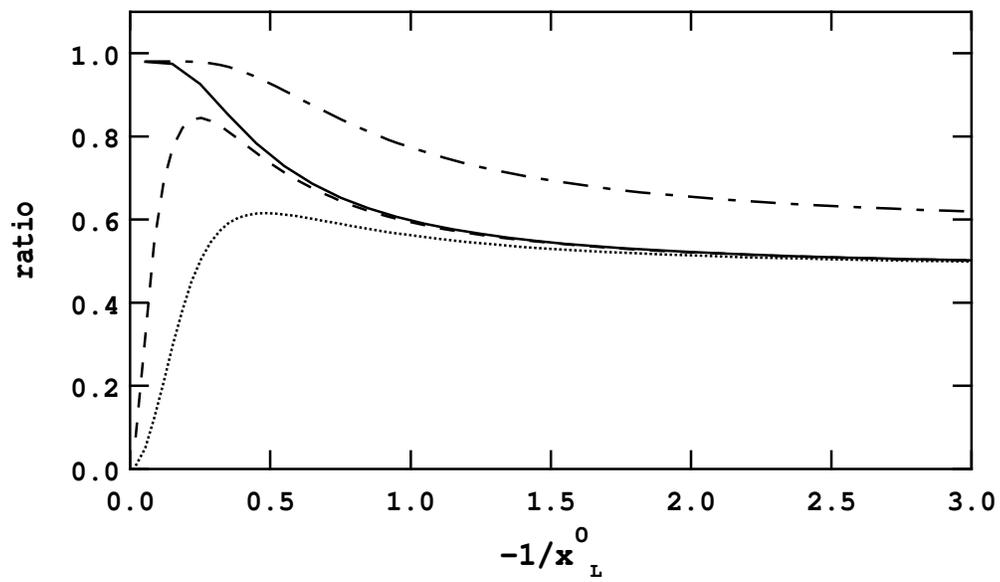
\begin{figure}
\caption{Dependence of the ratio 
${3\hbar}\dot{S}^{2}/\pi k_{B}^{2}\dot{E}$ on 
${-x^{0}_{L}}^{-1}=k_{B}T_{L}/\mu_{L}$ for $g=1$ with
$\mu_{L}=\mu_{R}$ (solid line), $\mu_{L}=1.01\mu_{R}$ (dashed line), 
and 
$\mu_{L}=1.1\mu_{R}$ (dotted line), and also for $g=1/2$ with 
$\mu_{L}=\mu_{R}$ (dot-dashed line). The parameter $x^{0}_{R}$ is chosen 
to be $x^{0}_{R}=100  
x^{0}_{L}$.}   
\label{Fig.1}
\end{figure}
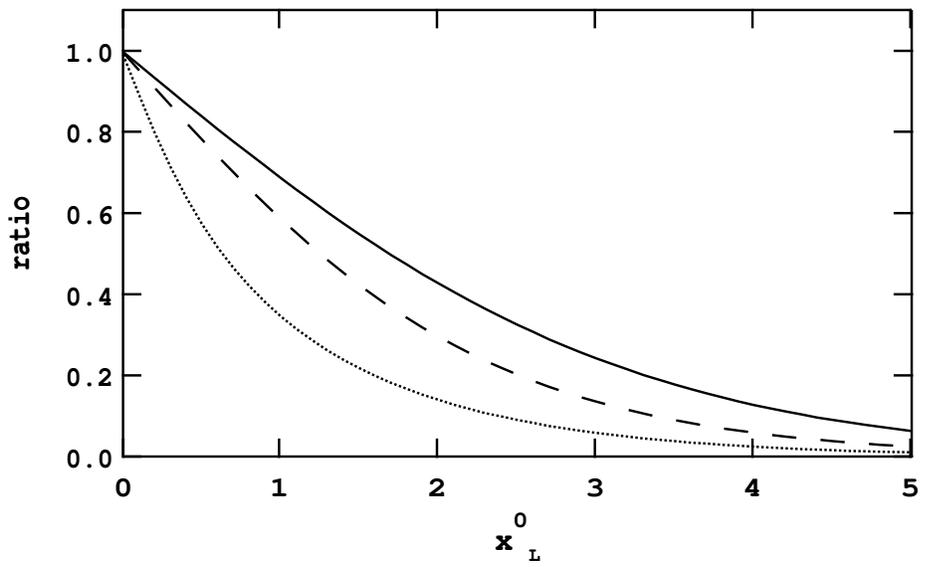
\begin{figure}
\caption{Dependence of the ratio 
${3\hbar}(T_{L}+T_{R})\dot{S}^{2}/[\pi k_{B}^{2}(T_{L}-T_{R})\dot{Q}]$ on
${x^{0}_{L}}=-\mu/k_{B}T_{L}$ for $g=0$ with $T_{R}=0.9 T_{L}$ 
(solid line), $T_{R}=0.5 T_{L}$ (dashed line), and $T_{R}=0.1 T_{L}$ 
(dotted line).}
\label{Fig.2}
\end{figure}
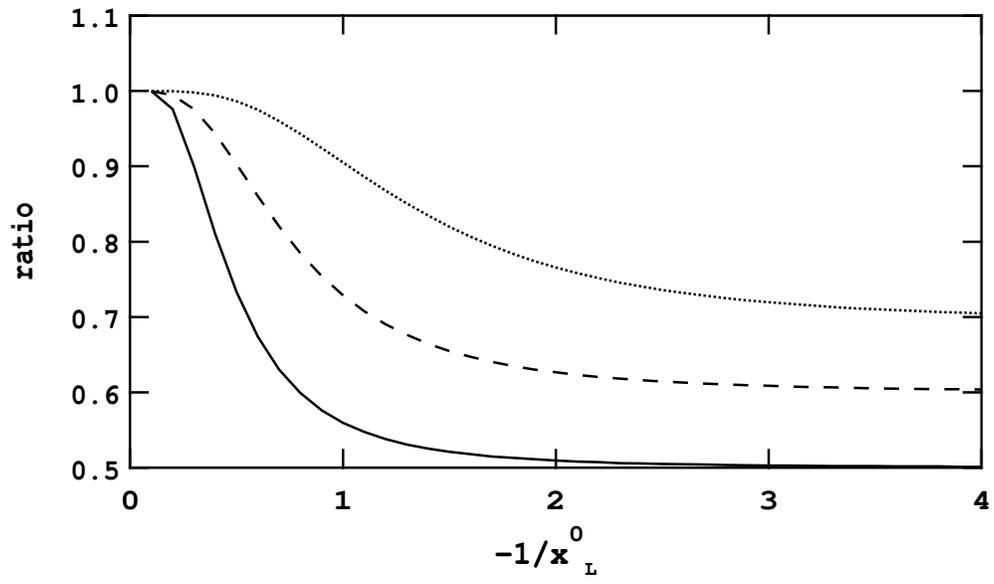
\begin{figure}
\caption{Dependence of the ratio 
${3\hbar}(T_{L}+T_{R})\dot{S}^{2}/[\pi k_{B}^{2}(T_{L}-T_{R})\dot{Q}]$ on
$-{x^{0}_{L}}^{-1}=k_{B}T_{L}/\mu$ for $g=1$  
(solid line), $g=1/2$ (dashed line), and $g=1/4$ 
(dotted line). The temperature $T_{R}$ is chosen 
to be $T_{R}=T_{L}/2$.}
\label{Fig.3}
\end{figure}
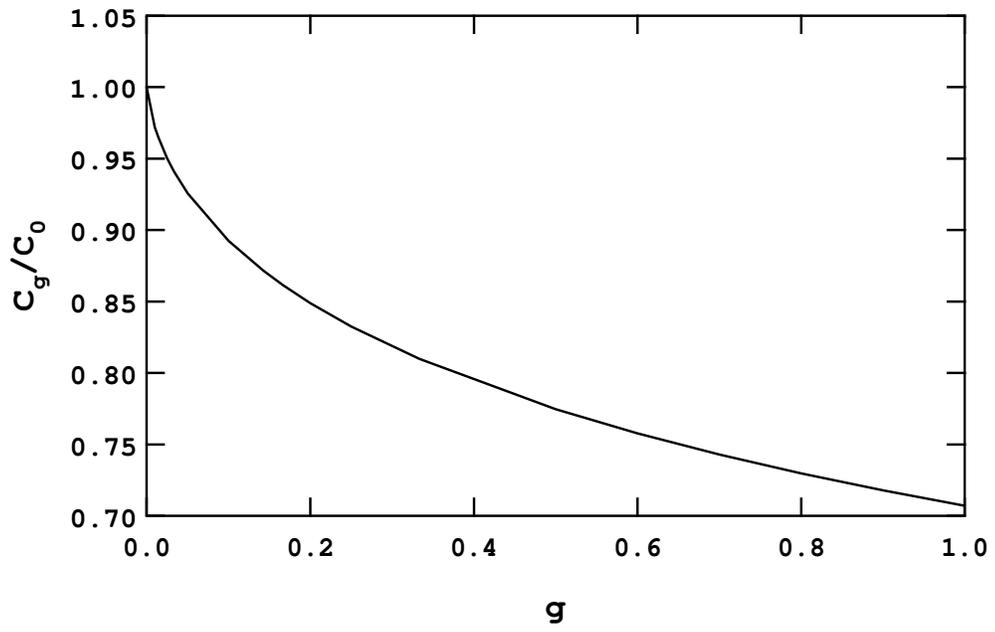
\begin{figure}
\caption{Dependence of the  optimum capacity ratio $C_{g}/C_{0}$ on 
the statistical parameter $g$. Note that only rational values of $g$ are 
physical.}
\label{Fig.4}
\end{figure}
\vfill
\eject

\mbox{\epsfig{file=universalfig1.EPSF, width=6in}}

\mbox{\epsfig{file=universalfig2.EPSF, width=6in}}

\mbox{\epsfig{file=universalfig3.EPSF, width=6in}}

\mbox{\epsfig{file=universalfig4.EPSF, width=6in}}

\end{document}